\documentclass[journal]{IEEEtran}

\usepackage{cite}

\usepackage{amssymb}

\usepackage[cmex10]{amsmath}
\usepackage{url}

\newcommand{\ceil}[1]{\lceil #1 \rceil}
\newcommand{\Fq}{\mathbb{F}_q}
\newcommand{\prob}[1]{P(#1)}
\newcommand{\Prob}[1]{P\left(#1\right)}

\newcommand{\ignore}[1]{}

\newtheorem{theorem}{Theorem}[section]
\newtheorem{lemma}[theorem]{Lemma}
\newtheorem{corollary}[theorem]{Corollary}
\newtheorem{definition}[theorem]{Definition}

\renewcommand{\paragraph}[1]{\vspace{0.15cm}\noindent {\bf #1}}

\begin{document}

\title{Analyzing Network Coding Gossip Made Easy\\ \large (Simpler Proofs for Stronger Results Even in Adversarial Dynamic Networks)}

\author{Bernhard~Haeupler
\thanks{B. Haeupler is with the Department
of Electrical Engineering and Computer Science, Massachusetts Institute of Technology, Cambridge,
MA, 02139 USA e-mail: (see http://people.csail.mit.edu/haeupler/).}

}

\maketitle

\begin{abstract}
We give a new technique to analyze the stopping time of gossip protocols that are based on random linear network coding (RLNC). Our analysis drastically simplifies, extends and strengthens previous results. We analyze RLNC gossip in a general framework for network and communication models that encompasses and unifies the models used previously in this context. We show, in most settings for the first time, that it converges with high probability in the information-theoretically optimal time. Most stopping times are of the form $O(k + T)$ where $k$ is the number of messages to be distributed and $T$ is the time it takes to disseminate one message. This means RLNC gossip achieves ``perfect pipelining''.

Our analysis directly extends to highly dynamic networks in which the topology can change completely at any time. This remains true even if the network dynamics are controlled by a fully adaptive adversary that knows the complete network state. Virtually nothing besides simple $O(kT)$ sequential flooding protocols was previously known for such a setting. 

While RLNC gossip works in this wide variety of networks its analysis remains the same and extremely simple. This contrasts with more complex proofs that were put forward to give less strong results for various special cases.
\end{abstract}


\section{Introduction}\label{sec:intro}
%
%
%
%
\IEEEPARstart{T}{his}
paper presents a new way to analyze gossip protocols based on random linear network coding that substantially simplifies, extends, and strengthens the results of previous work\cite{algebraicgossip-deb-med-choute-06-transinf,informationdissemination05,mosk2006information,vasudevan2009algebraic,borokhovich2010tight}. Gossip is a powerful tool to efficiently disseminate information. Its randomized nature is especially well-suited to work in unstructured networks with unknown, unstable or changing topologies. Because of this, gossip protocols have found a wide range of applications \cite{gkantsidis2005network,adhocrouting,demers1987epidemic,epidemicdatabase,kempe2004spatial}
and have been extensively studied over the past several decades \cite{hedetniemi88,topkis85,aspnes09,hromkovic96, kempe03, kempe02, karp00,minski-spreading,spreadingwithconductance}.

Recently, gossip protocols based on random linear network coding (RLNC)~\cite{ahlswede2000network,li2003linear,ho2003benefits} have been suggested ~\cite{algebraicgossip-deb-medard04-allerton} to cope with the additional complexities that arise when multiple messages are to be distributed in parallel. RLNC gossip has been adopted in many practical implementations \cite{gkantsidis2005network,katti2005importance,chou2003practical,efficientbroadcastingusingNC,katti2008xors,fragouli2006network} and has performed extremely well in practice. 

These successes stand in contrast to how little RLNC gossip is understood theoretically. 
Since its initial analysis on the complete graph \cite{algebraicgossip-deb-medard04-allerton,algebraicgossip-deb-med-choute-06-transinf, informationdissemination05}, several papers~\cite{vasudevan2009algebraic,borokhovich2010tight,mosk2006information} have tried to give good upper bounds on the stopping time of RLNC gossip in more general topologies. However, none of them address the case of unstable or changing topologies, and, even with the restriction to static networks, the guarantees are far from being general or tight on most graphs. In addition, all existing proofs are quite involved and do not seem to generalize easily.  

\paragraph{Our Results}

This paper has two main contributions.  The first is a new analysis technique that is both simpler and more powerful than previous approaches. Our technique relates the stopping time for $k$ messages to the much easier to analyze time $T$ needed to disseminate a single message. For the first time, and in practically all settings, this technique shows that RLNC gossip achieves perfect pipelining, i.e., it disseminates $k$ messages in order optimal $O(T + k)$ time. Our results match, and in most cases improve, all previously known bounds and apply to much more general models. To formalize this, we give a general framework for network and communication models that encompasses and unifies the models suggested in the literature so far. We give concrete results for several instantiations of this framework and give more detailed comparisons with previous results in each section separately.
 
As a second major contribution, our framework extends all models to
(highly) dynamic networks in which the topology is allowed to
completely change at any time. All of our results hold in these
networks even if the network dynamics are controlled by a fully
adaptive adversary that decides the topology at each time based on the
complete network state as well as all previously used
randomness. Virtually nothing, besides simple sequential flooding
protocols~\cite{KLO}, was previously known in such truly pessimistic
network dynamics. Having optimal ``perfectly pipelined'' stopping times in worst-case adaptive dynamic networks is among the strongest stability guarantees for RLNC gossip that one might hope for. To this end, our results are the first that formally explain RLNC gossip performance in the dynamic environments it is used in and was designed for. While the algorithm works in this wide variety of settings, our analysis remains mostly the same and extremely simple, in contrast with complex proofs that were previously put forward for the static setting.

\section{Background and Related Work}

Gossip is the process of spreading information via a randomized flooding procedure to all nodes in an unstructured network. It stands in contrast to structured multi-cast in which information is distributed via an explicitly built and maintained structure (e.g. spanning tree). While structured multi-cast can often guarantee optimal use of the limited communication resources it relies heavily on having a know and stable network topology and fails in distributed or uncoordinated settings. Gossip protocols were designed to overcome this problem. By flooding information in a randomized fashion they guarantee to deliver messages with high probability to all nodes with little communication overhead. This stability and distributed nature of gossip makes it an important tool for collaborative content distribution, peer-to-peer networks, sensor networks, ad-hoc networks and wireless networks and literature applying gossip in many areas and for many purposes is vast (e.g. \cite{gkantsidis2005network,adhocrouting,demers1987epidemic,epidemicdatabase,kempe2004spatial}). 

The gossip spreading of both a single message and multiple messages \cite{hedetniemi88,topkis85,aspnes09,hromkovic96, kempe03, kempe02, karp00,minski-spreading,spreadingwithconductance} has been intensely studied. The spreading of one message often follows a comparatively simple epidemic random process in which the message is flooded to a randomly chosen subset of neighbors. Spreading multiple messages in parallel is significantly more complicated because nodes need to select which information to forward. The main problem in this context is that widely spread messages get forwarded more often and quickly outnumber rarer messages. In many cases the slow spread of the rare messages dominates the time needed until all nodes know every message. 

A powerful and elegant way to avoid this and similar problems is the use of network coding techniques. Network coding as introduced by the seminal work of Ahlswede, Cai, Li and Yeung \cite{ahlswede2000network} breaks with the traditional concept that information is transported by the network as an unchanged entity. Ahlswede at al. show that in many multi-cast scenarios the optimal communication bandwidth can be achieved if and only if intermediate nodes in the network code information together. Li, Yeung and Cai \cite{li2003linear} showed that for multi-cast it is enough if intermediate nodes use linear coding, i.e. computing linear combinations of messages. Following this Ho, Koetter, M\'{e}dard, Karger and Effros \cite{ho2003benefits} showed that the coefficients for these linear combinations need not be carefully chosen with regard to the network topology but that for any fixed network the use of random linear combinations works with high probability. 

The strong performance guarantees and the independence of the coding procedure from any global information about the network makes random linear network coding (RLNC) the perfect tool for spreading multiple messages. This was first observed and made formal by Deb and M\'{e}dard \cite{algebraicgossip-deb-medard04-allerton}. They show that using randomized gossip and RLNC in a complete network in which each of the nodes starts with one message all information can be spread to all nodes in linear time, beating all non-coding approaches. After the introduction of this protocol in \cite{algebraicgossip-deb-medard04-allerton} and its follow-up \cite{algebraicgossip-deb-med-choute-06-transinf, informationdissemination05} it was used in many applications \cite{katti2005importance,chou2003practical,efficientbroadcastingusingNC,katti2008xors,fragouli2006network}, most notably the Microsoft Secure Content Distribution (MSCD) or Avalanche System \cite{gkantsidis2005network}. There has also been more theoretical work \cite{vasudevan2009algebraic,borokhovich2010tight,mosk2006information} investigating the convergence time of the RLNC-algorithm on general static network topologies. We give a detailed description and comparison to these works in section \ref{sec:applications}.

\paragraph{Gossip in Dynamic Networks Models}

While previous work on RLNC gossip focused on static networks our analysis shows that it works equally well in a wide range of dynamic network topologies. This contributes to ongoing work on modeling dynamic networks and exploring ways to efficiently communicate over them. With more and more modern networks being highly dynamic this task has recently gained importance. The model for studying these networks is still in flux. 

Substantial work has been devoted to random connectivity models in which a particular graph suffers different random edge faults in each round \cite{afek95}, or in which each node is connected to other random nodes in each round. Other work, e.g. on population protocols (see \cite{aspnes09} for a recent survey) has been invested in studying networks that eventually stabilize. Other models \cite{afek87,awerbuch88,awerbuch92,dijkstra74} allows for worst-case changes in network connectivity to happen, but only at a slow pace with plenty of time for self-stabilization to adapt to the changes. Gossip \cite{karp00,kempe02,hromkovic96,mosk-aoyama06,kempe03} and broadcasting \cite{bar-yehuda92,clementi01,clementi09,baumann09} are among the most frequently considered primitives in these settings.

Recently, Kuhn, Lynch, and Oshman\cite{KLO} proposed a truly pessimal model of network connectivity: that an adaptive adversary chooses the network structure
in each round, subject only to the requirement that the network be connected in each round, and that nodes \emph{anonymously broadcast} some chosen message without  knowing who their current neighbors are. The strength of this model means that any algorithms that work in it will be broadly applicable to dynamic networks. Kuhn et al. give simple algorithms based on sequentially flooding messages through the network as a proof that computation is at least possible though with strong performance losses compared to static networks (even a simple consensus takes $O(n^2)$ rounds in which all $n$ nodes communicate) . 

Our network model framework adopts the pessimal dynamics of Kuhn et al.~\cite{KLO} and can be seen as extending the model to also include network topologies with different connectivities, asynchronous communication or non-broadcasting behavior. More importantly is that this paper shows that RLNC gossip remains highly efficient in these dynamic networks giving the first improvements over the simple flooding algorithms in \cite{KLO}.

\paragraph{Organization}

Section \ref{sec:algorithm} reviews the RLNC algorithm and Section \ref{sec:technique} gives our new analysis technique. In Section \ref{sec:model} we introduce the network model framework. Section \ref{sec:applications} shows how to apply our technique in various instantiations of this framework. Section \ref{sec:extensions} finally discusses several ways in which the intentionally simple proofs from Section \ref{sec:applications} can be extended or sharpened.

\section{The RLNC Algorithm} \label{sec:algorithm}

In this section, we give a brief description of the RLNC algorithm.
The algorithm is simple and completely independent of the network structure or communication protocol. Alternative descriptions of the same algorithm can be found in \cite{algebraicgossip-deb-medard04-allerton} or \cite{chou2003practical}.

The RLNC algorithm sends out packets in the form of vectors over a finite field $\Fq$, where $q$ is an arbitrary prime or prime power. We assume that there are $k$ messages, $\vec {m_1},\ldots,\vec {m_k}$, that are vectors from $\Fq^l$ of length $l$. Every packet that is sent around during the execution of the algorithm has the form $(\vec \mu,\vec m)$, where $\vec m = \sum_{i=1}^k \mu_i \vec {m_i} \in \Fq^l$ is a linear combination of the messages, and $\vec \mu = (\mu_1,\ldots,\mu_k) \in \Fq^k$ is the vector of the coefficients. If enough packets of this form are known to a node, i.e., the span of the coefficient vectors is the full space $\Fq^k$, Gaussian elimination can be used to reconstruct all messages. For this, only $k$ packets with linearly independent coefficient vectors are needed. Linearity furthermore guarantees that any ``new packet'' that is created by taking a linear combination of old packets has the same valid format. With this, it is easy to see that a node can produce any packet whose coefficient vector is spanned by the coefficient vectors of the packets it knows. The algorithm is now easily described:

Each node $v$ maintains a subspace $X_v$ that is the span of all packets known to it at the beginning and received so far. If $v$ does not know any messages at the beginning, then $X_v$ is initialized to contain only the zero vector. If $v$ knows some message(s) $\vec m_i$ at the beginning, $X_v$ is initialized to contain the packet $(\vec \mu,\vec {m_i})$ in which $\vec \mu$ is the $i^{\text{th}}$ standard basis vector. $X_v$ furthermore contains all linear combinations that complete the span of these packet(s). Whenever node $v$ sends out a packet, it chooses a uniformly random packet from $X_v$. At the end of each round, all received packets are added to $X_v$ and again the span is taken. If the subspace spanned by the coefficient vectors is the full space, a node decodes all messages. 

Throughout the rest of the paper we will solely concentrate on the ``spreading'' of the coefficient vectors; the linear combination of the messages implied by a coefficient vector $\vec \mu$ is always sent along with it. We therefore define $Y_u$ to be only the coefficient part of $X_u$, i.e., the projection onto the first $k$ components.


{\bfseries Remark:} The parameter $q$ is used to trade of a faster running time versus bandwidth. While a larger $q$ can lead to faster convergence it increases communication overhead by increasing the size of the $k$ $(\log q)$-size RLNC-coefficients. In contrast to some of the related papers all results in this paper hold for arbitrary choices of $q$. For simplicity we will often restrict ourself to $q=2$. Note that this is the hardest case for running time considerations and it can be safely assumed that convergence times for larger $q$ will only be better. The case $q=2$ is furthermore interesting because it leads to the minimal RLNC-coefficients overhead and allows the use of simple XORs as a basic arithmetic operation.

\section{Our Technique} \label{sec:technique}

\subsection{Previous Approaches}

When analyzing the RLNC algorithm presented in Section \ref{sec:algorithm}, Sub and M\'{e}dard \cite{algebraicgossip-deb-medard04-allerton} were the first to use the notion of dimensionality of the subspaces $Y_v$ as a measure of progress. They made the observation that a node $u$ can, and most likely will, transmit new information to a node $v$, and thus increase the dimension of $Y_v$, whenever the subspace $Y_u$ is not already contained in $Y_v$. For this reason. they call such a node $u$ \emph{helpful} for $v$. It is easy to see that the vectors that do not extend the dimensionality of $v$, namely those in $Y_u \cap Y_v$, form a lower dimensional subspace in $Y_u$. This results in a success probability of at least $1 - 1/q$ if a random vector from $Y_u$ is chosen as a transmission. This fact and the notion of helpfulness is used as a crucial tool in all further RLNC proofs \cite{algebraicgossip-deb-med-choute-06-transinf,informationdissemination05,vasudevan2009algebraic,borokhovich2010tight,mosk2006information}.  

\subsection{Our Analysis Technique}

We argue that the right way to look at the spreading of information is to look at the orthogonal (dual) complement\footnote{While this section is self-contained Appendix \ref{app:orthogonal} offers additional information on orthogonal complements.} $Y_u^\perp$ of the coefficient subspaces $Y_u$. While the coefficient subspaces grow monotonically to the full space their orthogonal complement decreases monotonically to the empty span. To see how quickly this happens we first concentrate on one fixed (dual) vector $\vec \mu$, determine the time that is needed until it disappears from all subspaces $Y_u^\perp$ with high probability and than take a union bound over all those dual vectors. 

To formalize this we introduce the following crucial notion of knowing:

\vspace{0.17cm} \begin{definition}
A node $A$ knows about $\vec \mu \in \Fq$ if its coefficient subspace $Y_A$ is not orthogonal to $\vec \mu$, i.e., if there is a vector $\vec c \in Y_A$ with $<\vec c, \vec \mu> \neq 0$.
\end{definition} 
\vspace{0.1cm} 

Note that a node $A$ knowing a vector $\vec \mu$ does not imply $\vec \mu \in Y_A$ or anything about $A$ being able to decode a message associated with the coefficients $\vec \mu$. Knowing $\vec \mu$ only indicates that the node is not completely ignorant about the set of packets that have a coefficient vector orthogonal to $\vec \mu$.  Counterintuitively, because we are not working over a positive-definite inner-product space\footnotemark[1], it can even be that $\vec \mu \in Y_A$ but $A$ does not know $\vec \mu$.
For example, over $F_2^2$, if $Y_A$ is just (the span of) the vector $(1,1)$, then since $(1,1)$ over $F_2^2$ (has dot product 0 with itself mod 2), $A$ does not know $(1,1)$, even though $(1,1)\in Y_A$.  
 The next lemma proves the two facts that make this notion of knowledge so useful:

\vspace{0.17cm} \begin{lemma}\label{lem:knowledge-spreads}
If a node $A$ knows about a vector $\vec \mu$ and transmits a packet to node $B$ then $B$ knows about $\vec \mu$ afterwards with probability at least $1 - 1/q$. Furthermore if a node knows about all vectors in $\Fq^k$ then it is able to decode all $k$ messages. 
\end{lemma}
\vspace{0.1cm} \begin{IEEEproof}
Knowledge about a $\vec \mu$ essentially spreads with probability $1 - 1/q$ because the vectors in $Y_u$ that are perpendicular to $\vec \mu$ form a hyperplane in $Y_u$. For a complete and more elementary proof see Appendix \ref{app:proofs}.
\end{IEEEproof} \vspace{0.1cm}

With this, the spreading of knowledge for a vector $\vec \mu$ is a monotone increasing set growing process. It is usually relatively easy to understand this process and to determine its expected cover time $T$. Because the spreading process can be seen as a monotone Markov process, it is easy to prove that the cover time always has an exponentially decaying tail. In most cases this tail kicks in close to the expectation. This allows to pick a $t$ (usually $t = O(T + k)$) such that after $t$ time any vector in $\Fq^k$ has spread with probability $2^{-O(k)}$ and then take a union bound over all $q^k$ vectors to complete the proof that with high probability everything has spread. The following theorem summarizes this idea:

\vspace{0.17cm} \begin{theorem}\label{thm:reduction}
Fix a prime (power) $q \geq 2$, a probability $\delta > 0$ and an arbitrary network and communication model.\\
Suppose a single message is initiated at a node $v$ and then flooded through the network by the following faulty broadcast: In every round every node that knows the message and is supposed to communicate according to the communication model does forward the message with probability $1-1/q$ and remains silent otherwise. If for every node $v$ the probability that the message reaches all nodes after $t$ rounds is at least $1 - \delta q^{-k}$ then $k$ messages can be spread in the same model in time $t$ with probability $1 - \delta$ using the RLNC gossip protocol with field size $q$.
\end{theorem}
\vspace{0.1cm} \begin{IEEEproof}
This follows directly from the discussion above and Lemma \ref{lem:knowledge-spreads}. Initially every non-zero vector $\vec \mu \in \Fq^k$ is known to at least one node namely the one that knows about the $i$th message where $i$ is a non-zero component of $\vec \mu$. Whenever the network and communication model dictates that a node $A$ that knows $\vec \mu$ sends a message to a node $B$ Lemma \ref{lem:knowledge-spreads} shows that with probability $1 - 1/q$ the node $B$ afterwards knows $\vec \mu$. The spreading of each vector $\vec \mu$ therefore behaves like a faulty flooding process that floods $\vec \mu$ in every transmission with probability $1 - 1/q$. By assumption we have that after $t$ time steps every vector from $\Fq^k$ fails to spread to all nodes with probability at most $\delta q^{-k}$. Taking a union bound over all $q^{k}$ vectors gives the guarantee that the probability that after $t$ rounds all nodes know about all vectors is at least $1 - \delta$. According to Lemma \ref{lem:knowledge-spreads} all nodes can decode in this case and have learned the $k$ messages.
\end{IEEEproof}

\subsection{A Typical Template}\label{sec:simple-template}

Next we give a typical and easy way to apply Theorem \ref{thm:reduction}. We show that the cover time for one vector $\vec \mu$ is often dominated by a negative binomial distribution $NB(T,1-p)$, where $T$ is the expected coverage-time, and $p$ is a constant probability. Such a distribution has a strong enough tail to prove optimal $O(T + k)$ stopping times. In what follows we give a simple template to establish this:
 
What is needed for this template is a definition of a ``successful round'' such that at most $T$ such rounds are needed to spread a single vector $\vec \mu$ and such that a round is not a success with (say for now constant) probability at most $p$. The appropriate definition of success depends on the network model and is usually centered around its expansion, cuts, or diameter which determine how many additional nodes come to know about the vector in a ``good round''. 
Since nodes do not forget any information this spreading process is monotone and no progress gets lost in a bad round. Thus if the knowledge about $\vec \mu$ has not spread after $t = c(k + T  + \log \delta)$ steps, then there were at least $c(k+T + \log \delta)-T>(c-1)(k+T + \log \delta)$ failures, whereas one would only expect $p c (k + T)$. If we choose the constant $c$ large enough, a Chernoff bound or even simpler methods can now show that the probability for this to happen is at most $2^{-O(k + T + \log \delta)}$. This is small enough that, after a union bound over all $q^k$ vectors (e.g. for $q=2$), the probability that all $k$ messages have not spread is at most $\delta$. This simple template often applies  directly and leads to simple proofs of expected and high probability converges times of $O(k + T)$ that are often already order optimal. Even when not stated explicitly, all of our results hold furthermore with high probability.  In particular as shown here, an optimal additive $\Theta(\log \delta^{-1})$ additional rounds typically suffice to obtain a $1 - \delta$ success probability for any $\delta > 0$.

\section{Network Model and Communication Framework}\label{sec:model}

In this section, we elaborate on our network model framework that encompasses and extends the models suggested in the literature so far. The models and the results are very stable and can easily be extended further. We chose the following description as a trade-off between simplicity and generality.

\paragraph{The Network}

We consider networks that consist of $n$ nodes. A network is specified by a (directed) graph $G(t)$ on these nodes for every time $t$. Edges in $G(t)$ are links and present potential communication connections between two nodes in round $t$. We will usually assume that the network has, at all times, certain connectivity properties and will express the stopping time in terms of these parameters. (See also Section \ref{sec:modelextensions}.)

\paragraph{(Adversarial) Dynamics}

In all previous papers that analyzed the RLNC algorithm, the network topology was assumed to be \emph{static}, i.e., $\forall t: G(t)= G$. As discussed in the introduction, we allow the network topology to change completely from round to round and allow a fully adaptive adversary to choose the network. Because we are dealing with randomized protocols, we have to specify precisely what the adversary is allowed to adapt to. In our models (similar to \cite{KLO}) an \emph{adaptive adversary} gets to know the complete network state and all previously used randomness when choosing the topology. After that, independent randomness is used to determine the communication behavior and the messages of the nodes on this topology. 

\paragraph{The Goal: Gossip}

Distributed over the network are $k$ messages numbered $1,\ldots,k$ each known to at least one node. Throughout this paper, we assume a worst-case starting configuration for all messages including the case in which all messages are exclusively known to only one node (see also Section \ref{sec:mixed-initial-state}). The goal of gossip protocols is to make all messages known to all nodes in the network using as little time as possible (in expectation and with high probability)

\paragraph{Communication}

Nodes communicate along links with each other during transactions that are atomic in time. In each round, one packet is transmitted over a link if this link is activated in this round. From the view of a node, there are four commonly considered types of connections. Either a node sends to all its neighbors, which is usually referred to as BROADCAST, or it establishes a connection to one (e.g. uniformly random) neighbor and sends (PUSH) or receives (PULL) a message or both (EXCHANGE). In all cases, the packet is chosen without the sender knowing which node(s) will receive it.

\paragraph{Message and Packet Size}

As described in Section \ref{sec:algorithm} we assume that all messages and packets have the same size, and that a packet exactly contains one encoded message and its RLNC-coefficients. Note that the restriction on the message size is without loss of generality, since one can always cut a big message into multiple messages that fit into a packet. We also assume that the message size is large enough that the size of the RLNC-coefficients that are sent along is negligible. This assumption was made by all previous work and is justified by simulations and implementations in which the overhead is only a small fraction (e.g. $< 1\%$ \cite{algebraicgossip-deb-medard04-allerton}) of the packet size.

\paragraph{Synchronous versus Asynchronous Communication}

We consider two types of timing models. In the synchronous case, all nodes get activated at the same time and choose their messages independently, and messages get delivered according to the current network $G(t)$ and who sends and receives from whom. Note that this model is inherently discrete, and we assume that $t=1,2,\ldots$ are the times when nodes communicate.  We discuss this model in Section~\ref{sec:randomphonecall}.  For the asynchronous case, we assume that every node communication is triggered independently by a Poisson clock. This means that (with probability one) at any time only one node sends its message. This model can be directly translated into a discrete time model that defines round $i$ as the $i^{\text{th}}$ time such a communication takes place. 
The model considered in the literature so far assumes that every node is activated uniformly at random to communicate and then chooses a uniformly random neighbor for a PUSH, PULL or EXCHANGE. They also scale the time in the asynchronous model by a factor of $1/n$ so that each node gets activated once per time unit in expectation. We do not assume uniformity in either of the two distributions, and we present results for this more general model in Section \ref{sec:asynchsingle}.

\section{Applications and Results}\label{sec:applications}

In this section we take the models from Section \ref{sec:model} and describe the results that can be obtained for them using our analysis technique. There is a section for each different kind of communication model. We start with the Random Phone Call Model \cite{algebraicgossip-deb-medard04-allerton} that introduced RLNC-gossip. We than cover the extensions to arbitrary underlying network topologies as considered by  \cite{vasudevan2009algebraic,borokhovich2010tight,mosk2006information}. Section \ref{sec:asynchsingle} proves stopping times for a communication model that encompasses all former asynchronous communication protocols (PUSH, PULL, EXCHANGE, \ldots). For this model we answer a question of \cite{borokhovich2010tight} and show that a simple min-cut quantity exactly captures the behavior of gossip of $n$ messages. Lastly in Section \ref{sec:broadcast} we give the first bounds for the performance of synchronous and asynchronous BROADCAST in general networks. In this section we concentrate on showing only simple proofs that solely use the template from Section \ref{sec:simple-template}. In Section \ref{sec:extensions}, we revisit the models covered here and show some proof extensions.

\subsection{Random Phone Call Model and Gossip Mongering}\label{sec:randomphonecall}

In this section, we consider the work of Deb and M\'{e}dard~\cite{algebraicgossip-deb-medard04-allerton} and its follow-up~\cite{informationdissemination05,algebraicgossip-deb-med-choute-06-transinf} and show how to simplify and improve the analysis. The papers use a fairly simple model from our framework, namely the synchronous PUSH or PULL model on the complete graph, i.e., $G(t) = K_n$. This means in each round each node picks a random other node to exchange information with. This model is also known as the random phone call model and was introduced by \cite{demers1987epidemic}. It is shown~\cite{algebraicgossip-deb-medard04-allerton} that it is possible in this model  to spread $k=\Theta(n)$ messages in $O(n)$ time if $q = n$. This beats the $O(n \log n)$ time of $n$ sequential $O(\log n)$-phases of flooding just one message. The follow-up papers\cite{informationdissemination05,algebraicgossip-deb-med-choute-06-transinf} generalize this result to smaller number of messages $k$ and allow $q$ to be as small as $k$. They show that the running time of the algorithm is $t = O(k + \sqrt{k} \log k \log n)$, i.e., order optimal as long as $k \geq \log^3 n$. In order to prove this result, they have to assume that each node knows initially only one message and that initially the messages are equally spread. 
Even with these assumptions the analysis is long and complicated and the authors state themselves in their abstract that ``While the asymptotic results might sound believable, owing to the distributed nature of the system, a rigorous derivation poses quite a few technical challenges and requires careful modeling and analysis of an appropriate time-varying Bernoulli process.''

Our next lemma shows that RLNC gossip actually always finishes with high probability in order optimal stopping time $O(k + \log n)$. Our analysis is much simpler and has many further advantages: It holds for all choices of $k$ and allows $q$ to be as small as $2$. Our proof does also not rely on any assumptions on the initial message distribution. We show in Section~\ref{sec:exact-dependence-k} that the well-mixed initial state assumed in \cite{algebraicgossip-deb-medard04-allerton, informationdissemination05,algebraicgossip-deb-med-choute-06-transinf} actually provably speeds up the convergence compared to the worst-cast distribution for which our result holds. Our proof furthermore gives a success probability of $1 - 2^{t}$ if the algorithm runs for $O(t)$ time. In the setting of \cite{algebraicgossip-deb-medard04-allerton} with $k=n$, this is $1 - 2^{-n}$ instead of the $1-1/n$ stated there. Lastly it is interesting to note that previous general approaches~\cite{borokhovich2010tight,mosk2006information} are unable to prove any running time that beats the simple non-coding non-gossiping $O(n \log n)$ sequential flooding approach when applied to the complete graph/network. 

\vspace{0.17cm} \begin{lemma}\label{lem:randomphonecall}
The RLNC gossip in the random phone call model with $q=2$ spreads $k$ messages with high probability in exactly $\Theta(k + \log n)$ time. This holds independently from the initial distribution of the messages and of the communication model (e.g. PUSH, PULL, EXCHANGE). 
\end{lemma}

\subsection{Asynchronous single transfer protocols} \label{sec:asynchsingle}
After the helpfulness of RLNC gossip was established for the complete graph by \cite{algebraicgossip-deb-medard04-allerton}, the papers \cite{mosk2006information},\cite{vasudevan2009algebraic} and \cite{borokhovich2010tight} generalized it to general static topologies and consider asynchronous and synchronous PUSH, PULL and EXCHANGE gossip. In this section we first review the previous results and than show how to improve over them giving an exact characterization of the stopping time or RLNC gossip for $k=n$ messages using the template of Section \ref{sec:simple-template}.

The paper ``Information Dissemination via Network Coding''\cite{mosk2006information} by Mosk-Aoyama and Shah was the first to consider general topologies. They consider a similarly general version of the synchronous and asynchronous gossip as presented here and analyze the stopping times for $k=n$ in dependence on the conductance. Their analysis implies that with high probability $O(n \log n)$ phases of $n$ asynchronous rounds suffice for the complete graph and constant degree expanders and $O(n^2)$ such phases for the ring-graph. While the analysis is very interesting, these results do not beat the simple (non-coding) sequential flooding protocol and the stopping time of the ring-graph and many other graphs is even off by a factor of $n$. Their running times for the synchronous model are similar but lose another $\log n$-factor. Their dependence is on the success probability $1 - \delta$ is furthermore multiplicative in $\log \delta^{-1}$ because it stems from a standard probability amplification argument. 

Two recent papers~\cite{vasudevan2009algebraic,borokhovich2010tight} analyzed RLNC gossip using two completely different approaches. The second~\cite{borokhovich2010tight} points out that the analysis of the first~\cite{vasudevan2009algebraic} is flawed and prove that the asynchronous RLNC gossip on a network with maximum degree $\Delta$ takes with high probability $O(\Delta n)$ time. Their proof uses an interesting reduction to networks of queues and applies Jackson's theorem. They also give a tight analysis and lower bounds for a few special graphs with interesting behavior (see below). While their analysis is exact for few selected graphs the analysis is far from tight and in most graphs the maximum degree has nothing to do with the stopping time of RLNC gossip. The major question asked in \cite{borokhovich2010tight} is to find a characterizing property of the graph that determines the stopping time. 

We give exactly such a characterization for the asynchronous case with $k=n$ assuming a worst-cast message initialization. The model we use is a generalization of the classical PUSH, PULL and EXCHANGE model: We allow the topology in every round to be specified by a graph with directed and/or undirected edges and a probability weight $p_e$ on every edge $e$, such that the sum over all edges is at most 1. In every round each edge gets exclusively selected with probability $p_e$, i.e., in each round at most one edge gets selected. If the edge is undirected an EXCHANGE is performed and if a directed edge gets activated a packet is delivered in the direction of the edge. Note that this model is a generalization of the ``classical'' communication models. To obtain the probability graph from the undirected network with PUSH or PULL one just has to replace every undirected edge $\{u,v\}$ by two directed edges with probability weight $\frac{1}{n \Delta_u}$ and $\frac{1}{n \Delta_v}$ where $\Delta_u$ and $\Delta_v$ are the degrees of $u$ and $v$ respectively. To obtain the EXCHANGE protocol each undirected edge $\{u,v\}$ simply has the probability weight $\frac{1}{n \Delta_u}+\frac{1}{n \Delta_v}$.

Given such a network graph $G$ with probability weights $p_e$ we define the min-cut $\gamma(G)$ as:
$$\gamma(G) = \min_{\emptyset \neq S \subset V} \sum_{e \in \Gamma_G^+(S)} p_e$$
where $\Gamma_G^+(S)$ are all edges leaving a non-empty vertex-subset $S \subset V$ in $G$. The next two lemmas show that this quantity exactly captures how long RLNC gossip for $n$ messages takes.

\vspace{0.17cm} \begin{lemma}\label{lem:cut-asynch-single}
If for every time $t$ the min-cut of $G(t)$ is at least $\gamma$ then the asynchronous single transfer algorithm with $q=2$ spreads $n$ messages with probability at least $1-2^{-n}$ in  $O(\frac{n}{\gamma})$ time. 
\end{lemma} \vspace{0.1cm}

The next lemma proves that $O(\frac{n}{\gamma})$ is optimal.

\vspace{0.17cm} \begin{lemma}\label{lem:lowerbound-asynch-single}
With high probability, the asynchronous single transfer algorithm takes at least $\Omega(\frac{k}{\gamma})$ rounds to spread $k$ messages if it is used on \underline{any} fixed graph $G$ with (min-)cut $\gamma$ on which at least $\Theta(k)$ messages are initialized inside this cut. 
\end{lemma} \vspace{0.1cm}

Applying Lemma \ref{lem:cut-asynch-single} to the standard PUSH/PULL model gives a $O(n^2 \Delta)$ stopping time for any dynamic graph whose maximum degree is bounded by $\Delta$, which is the main result of \cite{borokhovich2010tight}. It also gives $O(n^2)$ for the complete graph (instead of the worst case $O(n^3)$ of \cite{borokhovich2010tight}) and nicely explains the behavior of the barbel graph and the extended barbel-graph that were considered by \cite{borokhovich2010tight}. The proof of Lemma \ref{lem:cut-asynch-single} can furthermore easily be extended to show that the dependency on the success probability is only logarithmic and additive in contrast to the previous work \cite{mosk2006information,borokhovich2010tight}.

\subsection{BROADCAST}\label{sec:broadcast}

In this section we give convergence results for synchronous and asynchronous BROADCAST gossip in arbitrary dynamic networks. These are to our knowledge the first results for the RLNC algorithm in such a setting. We think the results in this section are of particular interest for highly dynamic networks. The reason for this is that many of the highly unstable or dynamic networks that occur in practice like ad-hoc-, vehicular- or sensor-networks are wireless and thus have inherent broadcasting behavior.

To fix a model we first consider the simple synchronous broadcast model. We assume without loss of generality that the network graph $G$ is directed because any undirected edge can be replaced by its two anti-parallel directed edges. Having wireless networks in mind we also assume that in each round each nodes computes only one packet that is then send out to all neighbors. Our results also hold for the less realistic model where a node sends out a different packet to each neighbor. 

The parameter that governs the time to spread one message in a static setting is (not surprisingly) the diameter $D$ and it is easy to prove $\Theta(D + k)$ stopping times for $k$ messages using our technique. In a dynamic setting this is not true. Even for just one message, an adaptive adversary can, for example, always connect both the set of nodes that know about it and the set of nodes that do not know about it to a clique and connect the two cliques by one edge. Even though the graph $G(t)$ has diameter $2$ at all times, it  clearly takes at least $n$ rounds to spread one message. 
In order to prove stopping times in the adaptive adversaries model we switch to a parameter that indirectly gives a good upper-bound on the diameter for many graphs. The parameter we use is the isoperimetric number $h(G)$, which is defined as follows:
$$h(G) := \min_{S \subseteq V} \frac{|\Gamma_G^+(S)|}{\min(|\overline{S}|,|S|)},$$
where $\Gamma_G^+(S)$ are the nodes in $G$ outside of the subset $S$ that are in the directed neighborhood of $S$.

To give a few example values: for disconnected graphs $h(G)$ is zero and for connected graphs it ranges between $1$ and $\frac{2}{n}$; for a $k$-vertex-connected graph $G$ we have $h(G) = \Omega(\frac{k}{n})$ and $h(G) = \Theta(1)$ holds if and only if $G$ is a vertex-expander (or a complete graph). 

We are going to show that the expected time for one message to be broadcasted is at most $T = \frac{\log (n h(G))}{h(G)}$. This is $O(n)$ for a line and $O(\log n)$ for any vertex-expander. Our bound is tight in the sense that for any value $h$ with $1 \geq h \geq \frac{2}{n}$ there is a static graph $G$ that has diameter at least $O(T)$ and isoperimetric number $h(G) = \Theta(h)$. 
Having an upper bound on the time $T$ it takes to spread one message we again prove an perfectly pipelined time of $O(T + k)$ for $k$ messages:

\vspace{0.17cm} \begin{lemma}\label{lem:synchbroadcast}
The synchronous broadcast gossip protocol takes with high probability at most $O(\frac{\log (n h)}{h} + k)$ rounds to spread $k$ messages as long as the isoperimetric number of the graph $G(t)$ is at least $h$ at every time $t$.
\end{lemma} \vspace{0.1cm}

A similar result to Lemma \ref{lem:synchbroadcast} can be proven for the asynchronous BROADCAST model in which at every round each node gets selected uniformly independently at random (i.e. with probability $\frac{1}{n}$) to broadcast its packet to its neighbors:

\vspace{0.17cm} \begin{lemma} \label{lem:asynchbroadcast}
The asynchronous broadcast gossip protocol takes with high probability at most $O(n \cdot(\frac{\log (n h)}{h} + k))$ rounds to spread $k$ messages as long as the isoperimetric number of the graph $G(t)$ is at least $h$ at any time $t$.
\end{lemma}

\section{Extensions}\label{sec:extensions}

In this section we discuss how the simple proofs from Section \ref{sec:applications} that use only the template from Section \ref{sec:simple-template} can be extended to give more detailed or sharper bounds. 

\subsection{Exploiting a Well-Mixed Message Initialization} \label{sec:mixed-initial-state}

As stated in Section \ref{sec:model} we assume throughout the paper that $k$ messages are to be spread that are initially distributed in a worst-case fashion. All earlier papers restricted themselves to the easier special case that $k=n$ and that each node initially holds exactly one message \cite{borokhovich2010tight,mosk2006information}, or that $k$ is arbitrary but the network starts in a similarly well-mixed state in which each message is known by a different node and all messages are equally spread over the network \cite{algebraicgossip-deb-med-choute-06-transinf}. 
In many cases the worst-case and any well-mixed initialization take equally long to converge because the running time is lower bounded and bottlenecked by the flooding time $T$ for a single message or the time it takes for a node to receive at least $k$ packets. Nevertheless there are cases where a well-mixed initialization can drastically improve performance. 

Our proof technique explains this and we give a simple way to exploit assumptions about well-mixed initializations to prove stronger performance guarantees: If, e.g., each node initially holds exactly one of $k=n$ messages then most vectors $\vec \mu$ are already known to most nodes initially. More precisely exactly the $\binom{n}{i}(q-1)^i$ vectors with $i$ non-zero components are initially known to exactly $i$ nodes. With many vectors already widely spread initially the union bound over the failure probabilities for all vectors to spread after $t$ rounds can decrease significantly. Taking the different quantities and probabilities for nodes that are initially known to a certain number of nodes in account one can prove in theses cases that a smaller $t$ suffices.

One example for a mixed initialization being advantageous is discussed in the next Section \ref{sec:exact-dependence-k} and another one is the convergence time of the asynchronous PUSH and PULL protocol on the star-graph: For both PUSH and PULL the network induced by the star-graph has a min-cut of $1/n^2$ which leads according to Lemma \ref{lem:cut-asynch-single} and \ref{lem:lowerbound-asynch-single} to a stopping time of $\Theta(n^3)$ under a worst-case initialization. To lower bound the convergence time Lemma \ref{lem:lowerbound-asynch-single}, which relates the convergence time to the min-cut of the network graph, has to assume that at least a constant fraction of the messages are initialized inside a bad cut. For the ``classical'' initialization in which each node starts with exactly one message this is true for the PUSH model but not in the PULL model in which every bad cut only contains few messages. Indeed assuming a well-mixed initialization the PUSH protocol takes still $\Theta(n^3)$ time to converge while a much lower $\Theta(n^2 \log n)$ stopping time for the PULL model can be easily derived using our techniques.

\subsection{Exact Dependence on $k$ and Perfect Pipelining}\label{sec:exact-dependence-k}

In most (highly connected) networks the spreading time $T$ for one message is short and $O(k)$ becomes the dominant term in the order optimal $O(k + T)$-type upper bounds presented in this paper. So is, for example, $T = O(\log n)$ for most expanding networks. While it is clear that at least $k$ packets need to be received at each node it becomes an interesting question how large the constant factor hidden by the $O$-notation is. Differently stated, we ask how large the fraction of helpful or innovative packets received by a node is over the execution of the protocol. 

Determining and even more optimizing proofs to obtain such constants is usually a big hassle or even infeasible due to involved proofs. Simulation is therefore often used in practice to get a good estimation of the constants (e.g. \cite{algebraicgossip-deb-med-choute-06-transinf}). Our template from Section \ref{sec:simple-template} reduces the question for the stopping time of RLNC gossip to a simple standard question about tail bounds for negative binomial random variables. This makes it often possible to determine and prove (optimal) constants (and lower order terms). All that is needed is to replace the Chernoff bound in the template from Section \ref{sec:simple-template} by an argument that gives the correct base in the exponential tail-bound. In Section \ref{sec:tighter-tail} we give such a bound. We than exemplify then how to apply this bound by two examples: in Section \ref{sec:exact-broadcast} the synchronous BROADCAST gossip from Section \ref{sec:broadcast} and in Section \ref{sec:exact-rumormongering} the Rumor Mongering from Section \ref{sec:randomphonecall}. In both cases we can show that the constant in the dependency on $k$ is arbitrarily close to the absolutely optimal constant $1$, i.e. we can obtain a perfectly pipelined $t = k + O(T)$ stopping time.

\vspace{0.16cm} \subsubsection{A Tighter Tail Bound}\label{sec:tighter-tail}

The following simple lemma gives a stronger guarantee on the tail of a negative binomial random variable than the Chernoff bound used in the template from Section \ref{sec:simple-template}. The lemma proves that a constant factor away from the expectation the probability drops by a factor of $p$ with every additional trial instead of a constant factor drop that would be obtained by a standard Chernoff bound:

\vspace{0.17cm} \begin{lemma}\label{lem:tail}
The probability that after $t = k + O(T)$ independent trials there are less than $T$ successes is at most $p^k$ where $p$ is the failure probability (with $-\log p \geq \Omega(\log t)$).
\end{lemma}
\vspace{0.1cm}

If we apply this stronger tail bound in the template from Section \ref{sec:simple-template} we obtain the following corollary:

\vspace{0.17cm} \begin{corollary}\label{cor:tail}
Let $q \geq n^{\Omega(1)}$ and $T,k,d \leq n^{O(1)}$. If in order to spread any fixed coefficient vector $\mu$ only $T$ successful rounds are needed and 
if a round fails with probability at most $p$ 
then $k$ messages spread in $t = \frac{\log p}{\log q}k + O(T) + d$ rounds with probability at least $1 - p^d$. For $p=1/q$ this means a running time of $t = k + O(T)$ in expectation and with high probability.   
\end{corollary}

\vspace{0.16cm} \subsubsection{Perfect Pipelining of Synchronous Broadcast}\label{sec:exact-broadcast}

In this section we use the tighter tail bounds from the last Section \ref{sec:tighter-tail} to sharpen the bounds on the convergence time of the synchronous BROADCAST from Section \ref{sec:broadcast}:

\vspace{0.17cm} \begin{lemma}\label{lem:exact-synchbroadcast}
The synchronous broadcast gossip protocol takes with high probability at most $k + O(T)$ rounds to spread $k$ messages where $T = \frac{\log (n h)}{h}$ if the isoperimetric number of the graph $G(t)$ is at least $h$ at any time $t$. (and  $\log q = \Omega(\log n$))
\end{lemma}

\vspace{0.16cm} \subsubsection{Perfect Pipelining of Rumor Mongering}\label{sec:exact-rumormongering}

Another interesting case in which the exact dependence on the number of messages $k$ was considered is the Rumor Mongering process from Section \ref{sec:randomphonecall}. The authors of \cite{algebraicgossip-deb-med-choute-06-transinf} give a theoretical analysis in the regime $k > \log^2 n$ where the $O(k)$ term clearly dominates and prove an upper bound of $3.46k$ for the PUSH protocol and $5.96k$ for the PULL model. They also simulated the protocol and estimated the stopping time to be $1.5k + \log_2 n$. Both their analytic bounds and the simulation assume that messages start out in separate nodes and are equally spread over the network (see also Section \ref{sec:mixed-initial-state}). In this section we improve over these findings and show that the PULL model in this setting actually converges in $(1+o(1))k$ time for $k > \log^{1+o(1)} n$. Interestingly we also show that with a worst-cast initialization (see also Section \ref{sec:mixed-initial-state}) the PULL model does not achieve this convergence time but has a leading constant between $1.58$ and $1.83$:

Determining the correct constants for random communication protocols like the random phone call model is much more delicate than proving order optimal convergence times. The reason for this is that the union of random exchanges over many rounds almost surely form an expander while the graph in a single round is usually not even connected. This is the case for all of the presented random phone call models. While all these models are very stable order optimal one must be much more
careful to achieve and even more prove optimal $k(1 + o(1))$-type bounds for large $k$. We exemplify this by describing these concerns in detail for the PULL protocol:

The worst-case initialization for the PULL protocol is when all messages are initially known to only one node. In this case this node is not pulled at all in one round with probability $(1-1/n)^n \approx e^{-1} = 0.367879441$. In order to get pulled at least $k$ times it takes therefore in expectation at least $k / (1 - e^{-1}) = 1.58197671 k$ rounds. Thus for the case that only one node initially knows about all messages and if this node prepares a message in each round which it sends out to the nodes requesting it this is an information-theoretic lower bound on the number of rounds. A direct analysis of the protocol using Corollary \ref{cor:tail} for this case gives a constant of $\log(q) / \log( (1/e + (1-1/e)/q))$ which is $1.82462135 k$ for $q=2$. This can be improved if the start state is a bit more mixed, e.g., if each message is known to $i$ nodes initially. In this case the information-theoretical lower bound becomes $1/(1 - e^{-i}) k$ and our upper bound becomes $\log(q) / \log( (1/e^i + (1-1/e^i)/q))$ this means that for $i = \omega(1)$ our proof gives the optimal stopping time $t = k(1+o(1))$. Lemma \ref{lem:better-initialization} also shows a $(1+o(1)) k$ stopping time for the case where all messages are initiated at different nodes. This contrasts the upper bound of $5.96 k$ and the estimate of $1.5k$ of \cite{algebraicgossip-deb-med-choute-06-transinf} for this setting. More extensive simulation results than the ones in \cite{algebraicgossip-deb-med-choute-06-transinf} confirm that the constant for the dependency on $k$ should indeed be smaller than the projected $1.5 k$.

\vspace{0.17cm} \begin{lemma}\label{lem:better-initialization}
The RLNC algorithm in the random phone call PULL model even with $q=2$ spreads $k = \log^{1+o(1)} n$ messages with high probability in $(1+o(1)) k$ time if all messages are initially known to different nodes.  
\end{lemma}

\subsection{Asynchronous Single Transfer with small $k$}

Section \ref{sec:asynchsingle} proves convergence times for spreading $k=n$ messages using the asynchronous single transfer protocols. These bounds are tight and directly extend to a $\Theta(\frac{k}{\gamma})$ bound for $k = \Omega(n)$ messages. In what follows we want to generalize this to smaller number of messages and discuss the bounds that can be obtained using the technique from Section \ref{sec:technique}.

For small number of messages, e.g. $k=1$, the convergence time of RLNC single transfer gossip can be much faster than $O(\frac{n}{\gamma})$ but still be $\omega(\frac{k}{\gamma})$. This shows that the min-cut $\gamma$ is not the right quantity to look at in this scenario. Again, as in Section \ref{sec:broadcast}, conductance quantities capture much better how fast a small number of messages spreads. The quantity we consider is:
$$\lambda(G) = \min_{S \subset V} \frac{\sum_{e \in \mathbf{out}(S)} p_e}{\min(|S|,|\overline{S}|)}$$

The next lemma shows that it takes at most $T = O(\frac{\log n}{\lambda})$ time for one message to spread if the conductance is bounded by $\lambda$.

\vspace{0.17cm} \begin{theorem}\label{thm:one-message-asynch-single}
In the asynchronous single transfer model (with any $q$) it takes in expectation at most $T = O(\frac{\log n}{\lambda})$ time for one message to spread. 
\end{theorem}
\vspace{0.1cm} \begin{IEEEproof}
The probability that a set of nodes that know about the message grows from size $i<n$ to $i+1$ is at least $(1-1/q) (\lambda \min(k,n - k))$. It thus takes at least $\frac{1}{(1-1/q)\lambda}$ rounds in expectation for the first success, $\frac{1}{2(1-1/q)\lambda}$ rounds for the second success and in general $T = \sum_{i=1}^{n} \frac{1}{\min(i,n - i) (1-1/q) \lambda} < (1 + \frac{1}{q-1}) \frac{2}{\lambda} \log n = O(\frac{\log n}{\lambda})$ rounds in expectation for one message to spread. 
\end{IEEEproof}\vspace{0.1cm}

This is a tight bound for many regular graphs and gives e.g. a flooding time of $\Theta(n \log n)$ for the complete graph or any other regular expanders. It is clear that RLNC-gossip for any $k$ needs to take at least so much time. The other lower bound that kicks in for large enough $k$ is the $\Omega(\frac{k}{\gamma})$ lower bound from Lemma \ref{lem:lowerbound-asynch-single}. Similar to the results for the other models we want show that the total running time is essentially (up to at most a $\log n$ factor) either dominated by the $T = \frac{\log n}{\lambda}$ rounds to spread one message or for larger number of messages $k$ the $O(\frac{k}{\gamma})$ rounds coming from the communication lower bound that the $k$ messages have to cross the worst case cut. 

\vspace{0.17cm} \begin{lemma}\label{lem:exact-asynch-single}
Disseminating $k$ messages in the asynchronous single transfer model with $q=2$ takes with high probability at most $t = O(\frac{k}{\gamma} + \frac{\log^2 n}{\lambda})$ rounds if the graph $G$ as a min-cut of at most $\gamma$ and a conductance of at least $\lambda$ at all times $t$. 
\end{lemma}




\subsection{Weaker Requirements for Random Networks} \label{sec:modelextensions}

The idea behind proving performances in the rather strong adaptive adversary model introduced in this paper is that the guarantees directly extend to the widest possible range of dynamic networks including random models. Most of our proofs like the ones of Lemma \ref{lem:cut-asynch-single}, \ref{lem:synchbroadcast} or \ref{lem:asynchbroadcast} demand that the network graph $G(t)$ has a certain connectivity requirement at any time $t$. These requirements might be too strong especially for random network models. We discuss in the following how these requirements can be easily weakened in many ways:

The simple fact that no progress in the spreading of knowledge gets lost makes it easy to deal with the case that the connectivity fluctuates (e.g., randomly). Increasing the stopping time by a constant factor easily accounts for models in which the desired connectivity occurs only occasionally or with constant probability. Looking at the average connectivity is another possibility. It is furthermore not necessary to require the entire graph to be expanding on average but it suffices to demand that each subset expands with constant probability according to its size. This way convergence can be proven even for always disconnected graphs. Especially for random models it can also be helpful to consider the union of the network graphs of consecutive rounds, i.e. $G'(t) = {G(3t')} \cup {G(3t'+1)} \cup {G(3t'+2)}$. This gives for example directly valid upper bounds for the synchronous or asynchronous BROADCAST model. 

As a simple example for the usefulness of these approaches we discuss an alternative way to prove Lemma \ref{lem:randomphonecall} about the stopping time of the Rumor Mongering process: Instead of analyzing the Rumor Mongering as a synchronous protocol on the complete graph in which each node performs a PULL, PUSH or EXCHANGE one can alternatively see it as a synchronous BROADCAST (see Section \ref{sec:broadcast}) on a random network. The network graph $G(t)$ in this case is simply formed by a random directed in-edge, directed out-edge or undirected edge at each node depending on whether on looks at the PUSH, PULL or EXCHANGE model. The results from Lemma \ref{lem:synchbroadcast} or \ref{lem:synchbroadcast} will not directly give any bounds simply because the network graph $G(t)$ is with high probability disconnected. Using either of the two more advanced extensions solves this problem: with constant probability every set has a constant expansion; alternatively one can use that the union of a constant number of rounds, as described above, forms with an expander with high probability.

\section{Conclusions and Open Questions}

We have given a new technique to analyze the stopping times of RLNC-gossip that drastically simplifies, strengthens and extends previous results. Most notably all our results hold in highly dynamic networks that are controlled by a fully adaptive adversary. 

Theorem \ref{thm:reduction} gives a direct way to transfer results for the single-message flooding/gossip process to the multi-message RLNC-gossip if strong enough tail bounds are provided. One candidate for which this could work is, e.g., \cite{spreadingwithconductance} which can be interpreted as giving bounds on a synchronous single transfer gossip for one message.

This paper also gives evidence that in most network models RLNC-gossip achieves perfect pipelining, i.e. the bounds for disseminating $k$ messages have the form $O(k + T)$ where $T$ is the expected time to (faultily) flood one message. It is a very intriguing question under which general conditions on the network model one can prove this behavior. It is easy to see that the monotone set-growing process induced by the faulty flooding process of one message always exhibits a strong exponential tail as needed to apply Lemma \ref{thm:reduction}. This already implies asymptotic convergence times of the form $\frac{k}{\gamma}(1 + o(1))$ (see also Lemma \ref{lem:exact-asynch-single}) where $\gamma$ is the min-cut in the induced Markov-Chain, i.e. the minimal probability over all sets to inform another node within one round. The main question remaining is therefore to guarantee that this tail kicks in after $O(T)$ rounds.

\appendices

\section{Preliminaries: Orthogonal Dual Complement} \label{app:orthogonal} 

In this section we provide a few background facts in linear algebra on vector spaces without (positive-definite) inner product, especially the notions involved in orthogonality. Even so the Section \ref{sec:technique} is fully 
self-containing this section might be helpful in understanding the proofs. 

For a vector space $V$ the \emph{dual space} $V^*$ consists of all linear forms on $V$. For any subset $S \subseteq V$ the orthogonal (dual) complement $S \perp$ is defined as all elements from $V^*$ that disappear on $S$. It is easy to see that the orthogonal complement is a subspace in $V^*$ and has co-dimension equal to the dimension of the span of $S$ in $V$. The dual space $V^*$ is isomorphic to $V$ and in the case of $\Fq^k$ the dot-product $y \mapsto (x \mapsto <x,y>)$ is an isomorphism. Using this identification the orthogonal complement can also be defined as the space of all vectors that are perpendicular (i.e. having a zero dot-product) to all vectors in $S$. This is the standard definition of orthogonality and for inner-product spaces like $R^k$ it matches the geometrical notion of orthogonality. This is not true for $\Fq^k$ in which the dot-product is not positive definite. This leads to counter-intuitive situations, e.g. the vector $[1,1]$ is orthogonal to itself in $F_2^2$. But the fact remains that every subspace $S \subseteq \Fq^k$ can be assigned a orthogonal complement subspace $S^\perp$ with $\dim(S) + \dim(S^\perp) = k$ remains true and is the important notion used in Section \ref{sec:technique}.

\section{proofs}\label{app:proofs}

\begin{IEEEproof}[Proof of Lemma \ref{lem:knowledge-spreads}]
We give a more basic proof here: For this we define two vectors $\vec c_1,\vec c_2 \in \Fq^k$ as equivalent if $<\vec c_1,\vec \mu> = <\vec c_2,\vec \mu>$. This splits $Y_A$ in exactly $q$ equivalence classes of equal size. To see this note that, because $Y_A$ is a subspace, scalar-multiplication is a bijection between any two equivalence classes that correspond to a non-zero dot-product. By assumption $Y_A$ furthermore contains a vector $\vec c$ that has a non-zero dot-product with $\vec \mu$. This gives that $\vec \mu$-translation is a bijection between the zero dot-product equivalent class and another equivalence class.
 Thus with probability exactly $1 - 1/q$ a packet with coefficient vector from a non-zero equivalence class is chosen for transmission. In this case this coefficient vector gets added to $Y_B$ and the node $B$ now knows $\vec \mu$.

For the second claim we prove that any node $A$ that is not able to decode does not know about at least one vector $\vec \mu$: If $A$ can not decode than $Y_A$ is not the full space. 
Because $Y_A$ is a subspace it is lower-dimensional and we can use Gram-Schmidt to construct a orthogonal basis of $Y_A$ and a vector $\vec \mu$ that is orthogonal to $Y_A$. This vector $\vec \mu$ is then by definition not known to $A$, a contradiction. 
\end{IEEEproof}\vspace{0.1cm}

\vspace{0.1cm} \begin{IEEEproof}[Proof of Lemma \ref{lem:randomphonecall}]
For the lower bound we note that each node receives in expectation (and with high probability) only $\Theta(1)$ packets per round. Thus if in the beginning at least one node did not already know about a constant fraction of the messages, then the algorithm has to run for at least $\Omega(k)$ rounds. It is also clear that even one message takes in expectation $\Omega(\log n)$ time to spread to all nodes. This completes the lower bound. 

To prove the upper bound, we use the template from \ref{sec:simple-template}: 
%
For this we fix a coefficient vector $\vec \mu$ and define a round as successful if the number of nodes that know about it increases by at least a constant factor $\lambda > 1$ or if the number of nodes that do not know about $\vec \mu$ decreases by a factor of $\lambda$. There are at most $O(\log n)$ successful rounds needed until at least $n/2$ nodes know about $\vec \mu$ and at most another $O(\log n)$ successful rounds until all nodes know about $\vec \mu$. It remains to be shown that each round succeeds with constant probability. 

We first consider the PULL model. At first we have $i < n/2$ nodes that know about $\vec \mu$ and at least $n/2$ nodes pulling for it. Each of those nodes has a probability of $i/n$ to hit a knowing node. We expect a $i/n$ fraction of the ignorant nodes, i.e., at least $i/2$ nodes, to receive a message from a node that knows about $\vec \mu$. The independence of these successes and Lemma \ref{lem:knowledge-spreads} prove that with constant probability at least $\Omega(i)$ nodes learn about $\vec \mu$. Once there are at least $n/2$ nodes that know $\vec \mu$, each of the ignorant nodes pulls a packet from a knowing node with probability at least $1/2$. 

The proof for the PUSH model is similar. If there are $i < n/2$ nodes that know about $\vec \mu$ and push out a message, then there are at least $n/2$ ignorant nodes that each receive at least one message from one of the $i$ nodes with probability $1 - (1 - 1/n)^i$. It is not hard to see that, in total, $\Omega(i)$ ignorant nodes receive a message from a node that knows $\vec \mu$ with constant probability. Lemma \ref{lem:knowledge-spreads} now guarantees that, with constant probability, the number of ignorant nodes that learn $\vec \mu$ is only a small factor smaller. Once there are $n/2$ nodes knowing about $\vec \mu$ and each of these pushes out, each node that does not know $\vec \mu$ has a chance of $(1-1/n)^{n/2} = e^{-2}$ per round to receive a message from a node that knows $\vec \mu$. Applying Lemma \ref{lem:knowledge-spreads} again finishes the proof. 
\end{IEEEproof}\vspace{0.1cm}

\vspace{0.1cm} \begin{IEEEproof}[Proof of Lemma \ref{lem:cut-asynch-single}]
Our proof proceeds along the lines of the simple template from Section \ref{sec:simple-template} and concentrates on the spreading of one coefficient vector. We define a round as a success if and only if one more node learns about it. It is clear that exactly $n$ successes are needed. From the definition of $\gamma$ and Lemma \ref{lem:knowledge-spreads} follows that each round is successful with probability at least $\gamma(1 - 1/q)$. Thus if we run the protocol for $t = c(\frac{n}{(1 - 1/q)\gamma})$ rounds we expect at least $c n$ successes and by Chernoff bound the probability that we get less than $n$ is at most $2^{-O(n)}$. If we choose $c$ appropriately this is small enough to end up with $2^{-n}$ after taking the union bound over the $q^k = 2^n$ vectors.
\end{IEEEproof}\vspace{0.1cm}

\vspace{0.1cm} \begin{IEEEproof}[Proof of Lemma \ref{lem:lowerbound-asynch-single}]
In each round, at most one packet can cross the cut. For this to happen, an edge going out of the cut has to be selected and the probability for this is by definition exactly $\gamma$. In order to be able to decode the $k$ messages at least $\Theta(k)$ packets have to cross the cut each taking in expectation at least $O(\frac{1}{\gamma})$ rounds. It takes with high probability at least $\Omega(\frac{k}{\gamma})$ rounds until $\Theta(k)$ packets have crossed the cut. 
\end{IEEEproof}\vspace{0.1cm}

\vspace{0.1cm} \begin{IEEEproof}[Proof of Lemma \ref{lem:synchbroadcast}]
We use the simple template from Section \ref{sec:simple-template} and concentrate on the spreading of one coefficient vector $\vec \mu$. We define a round to be a success if and only if the number of nodes that know about $\vec \mu$ grows at least by a $\frac{h}{7}$ fraction or the number of nodes that do not know about $\vec \mu$ shrinks at least by the same factor.\\
We want to argue that at most $T = O(\frac{\log (n h)}{h} )$ successes are needed to spread $\vec \mu$ completely. Note that this is slightly better than the straight forward $(1 + \frac{h}{7})^T \geq n$ bound that would lead to $T = O(\frac{\log(n)}{h})$. The improvement comes from exploiting the fact that the number of nodes that learn is an integral quantity: In the first $\frac{7}{h}$ successful rounds at least one node learns about $\vec \mu$. The next $\frac{7}{2h}$ successful rounds at least $2$ nodes learn about $\vec \mu$ and the following $\frac{7}{3h}$ successful rounds it is $3$ new nodes and so on. There are $\frac{n}{2} \cdot \left(\frac{7}{h}\right)^{-1}$ such phases until at least $n/2$ nodes know about $\vec \mu$. The downward progression than follows by symmetry. The total number of successes sums up to:
$$T \leq 2 \frac{7}{h} \sum_{i = 1}^{O(n h)} \frac{1}{i} = O(\frac{\log n h}{h}).$$
To finish the proof we show that every round has a constant success probability. This follows from Lemma \ref{lem:knowledge-spreads} if for a success only one node is supposed to learn about $\vec \mu$. If at least $\ceil{i} \geq 2$ nodes are supposed to learn then by the definition of a success and of $h(G(t))$ there are $k \geq \ceil{7i} \geq 4 \ceil{i}$ nodes on the knowledge cut, i.e., at least $k$ nodes that do not know about $\vec \mu$ are connected to a node that knows about $\vec \mu$. We invoke Lemma \ref{lem:knowledge-spreads} again to see that each of these nodes fails to learn about $\vec \mu$ with probability at most $1/q \leq 1/2$. Finally Markov's inequality gives that the probability that more than $k - \ceil{i} \geq \frac{3}{4} k$ fail to learn is at most $2/3$. A round is therefore successful with probability at least $1/3$.
\end{IEEEproof}\vspace{0.1cm}

\vspace{0.1cm} \begin{IEEEproof}[Proof of Lemma \ref{lem:asynchbroadcast}]
The proof is nearly identical to the one of Lemma \ref{lem:synchbroadcast} but instead of defining a round as a success we define successes for phases of $n$ consecutive rounds. Using the same definition of success and following the same reasoning as before it is clear that at most $O(\frac{\log (n h)}{h})$ successful phases are needed. To finish the proof we have to show that every phase has a constant success probability. For this we note again that at least $k \geq 4 \ceil{i}$ nodes are on the knowledge-cut of $\vec \mu$ if $\ceil{i}$ nodes need to learn about $\vec \mu$. For each of these $4 \ceil{i}$ nodes the probability that no neighboring node that knows $\vec \mu$ is activated during $n$ rounds is at most $(1 - 1/n)^{n} = e^{-1}$. According to Lemma \ref{lem:knowledge-spreads} the probability for each of the $k$ nodes to fail to learn about $\vec \mu$ is thus at most $1 - (1 - 1/q)(1 - e^{-1}) < 0.7 < 3/4$. Markov's inequality again implies that the probability for a failed round in which more than $k-\ceil{i} \geq 3/4k$ fail is at most $0.7/0.75$.
\end{IEEEproof}\vspace{0.1cm}

\vspace{0.1cm} \begin{IEEEproof}[Proof of Lemma \ref{lem:tail}]
We pick $t = k - (T+1)\log t/\log p + T$ and have now that 
\[ p^k = p^{t-T} t^{T+1} > \sum_{i=t-T}^t \binom{t}{t-i} p^i (1-p)^{t-i} \]
which is exactly the probability for having at least $t-T$ failures in $t$ rounds.
\end{IEEEproof}\vspace{0.1cm}

\vspace{0.1cm} \begin{IEEEproof}[Proof of Corollary \ref{cor:tail}]
Follows directly by applying Theorem \ref{thm:reduction} according to the template in Section \ref{sec:simple-template} and the use of Lemma \ref{lem:tail} to get the right bound on the tail probability. 
\end{IEEEproof}\vspace{0.1cm}

\vspace{0.1cm} \begin{lemma}\label{lem:weighted-sum-bernoulli}
Let $X_1, X_2, \ldots, X_l$ be i.i.d. Bernoulli variables with probability $\prob{X_1 = 0} = p \leq \frac{1}{2}$. The probability that a positively weighted sum of the variables is at most $\frac{1}{4}$ its expectation is at most $p$:
$$\forall w_1,\ldots,w_l > 0:\ \ \prob{\sum_j w_j X_j \leq \frac{1}{4} (1 - p) \sum_j w_j} \leq p.$$
\end{lemma}
\vspace{0.1cm} \begin{IEEEproof}
We first scale the weights such that $\sum_j w_j = 1$ and than use the second moment method: 

\begin{multline*}
\ \Prob{\ \ \  \sum_j w_j \ \ \ \ \ \ \ X_j \, \ \ \ \ \ \ \ \ \ \ \ \ \ \ \ \ \      \leq \ \ \ \ \ \   \frac{1}{4}  (1 - p)      \ \ \ \  } \\
= \Prob{\ \ \  \sum_j w_j (1 - X_j) - p\sum_j w_j \ \ \ \    \geq 1 -           \frac{1}{4}  (1 - p) - p           } \\
= \Prob{\ \ \  \sum_j w_j (1 - X_j) - p\sum_j w_j  \ \ \ \   \geq \ \ \ \ \ \   \frac{3}{4}  (1 - p)      \ \ \ \  } \\ 
= \Prob{\left( \sum_j w_j (1 - X_j) - p\sum_j w_j\right)^2  \geq \ \ \ \ \ \   \frac{9}{16} (1 - p)^2    \ \    }\\
\end{multline*}

Now the left-hand side is the variance of a weighted sum of i.i.d. Bernoulli variables with probability $1-p$, and as such its expectation is exactly $\sum_j w_j^2 (1-p)p$. Using Markov's inequality on this expectation, we get that the probability we want to bound is at most:
\begin{align*}
\left(\sum_j w_j^2 (1-p)p\right) \left(\frac{9}{16} (1 - p)^2\right)^{-1} &= \frac{16}{9} \frac{p}{1-p} \sum_j w_j^2\\
&\leq \frac{16}{9} \ \ 2p \ \ 1/4 \ \leq p.
\end{align*}
The last transformation holds because $1 - p \geq 1/2$ and because we can assume that all weights are at most $1/4$. This is true because if there is a $w_i \geq 1/4$ then already $X_i=1$ leads to an outcome of at least $1/4$ the expectation and the probability for this to happen is $p$. 
\end{IEEEproof}\vspace{0.1cm}

\vspace{0.1cm} \begin{IEEEproof}[Proof of Lemma \ref{lem:exact-synchbroadcast}]
We modify the proof of Lemma \ref{lem:synchbroadcast} only in the way that we use the stronger tail bound from Corollary \ref{cor:tail} instead of the simpler template from Section \ref{sec:simple-template}. We keep the same definition of success but prove that the success probability of a round is at least $1/q$ instead of $1/4$ as in Lemma \ref{lem:synchbroadcast}:

If only one node is supposed to learn for a success this is again clear by Lemma \ref{lem:knowledge-spreads}. If at least $\ceil{i}$ nodes nodes are needed to a success we know also by the definition of a success that at least $4 \ceil{i}$ nodes that do not know about $\vec \mu$ are connected to a node that knows about it. We assign each ignorant node to exactly one node that knows about $\vec \mu$ breaking ties arbitrarily. Now according to Lemma \ref{lem:knowledge-spreads} with probability $1-1/q$ each such node independently sends out a message that is not perpendicular to $\vec \mu$ and all ignorant nodes that are connected to it learn $\vec \mu$. We can now directly apply Lemma \ref{lem:weighted-sum-bernoulli} and obtain that we indeed have a success probability of at least $1/q$ per round. This finishes the proof. 
\end{IEEEproof}\vspace{0.1cm}

\vspace{0.1cm} \begin{IEEEproof}[Proof of Lemma \ref{lem:better-initialization}]
We assume each message is initially known to exactly one node and all messages are known to different nodes. This implies that exactly the $\binom{k}{i}(q-1)^i$ vectors that have $i$ non-zero components are initially known to exactly $i$ nodes. We will prove that the running time $t > k + O(\log n) \log t$ suffices to spread all messages with probability at least $1 - n^{-\Omega(1)}$.

For this we pick a threshold $f = \omega(1)$ and first look at the $\sum_{i=1}^{f} \binom{k}{i}(q-1)^i \leq f k^f$ vectors that are known to at most $f$ nodes initially. From the proof of Lemma \ref{lem:randomphonecall} we know that after $t$ rounds each of these vectors has a probability of at most $2^{-O(t - O(\log n))}$ to not have spread completely. Choosing $t > k + O(\log n)$ therefore suffices easily to make the contribution of these vectors to the union bound at most $n^{-\Omega(1)}$.

Most of the $q^k$ vectors start initially known to at least $f$ nodes. For these vectors $\vec \mu$ we choose the same definition of success as in the proof of Lemma \ref{lem:randomphonecall}: A round is successful if the number of nodes that know about $\vec \mu$ increases by at least a constant factor $\lambda > 1$ or if the number of nodes that do not know about $\vec \mu$ decreases by a factor of $\lambda$. We will show that if we choose $\lambda$ small enough these vectors have a probability of $\frac{1}{q}$ to spread successfully in one round.

While with our initial analysis the start phase was the critical bottleneck we can show that the success probability for this phase can now even be pushed below $1/q$ by choosing $\lambda$ small enough. In the first phase we have $k < n/2$ nodes that know $\vec \mu$ and at least $n/2$ nodes that are pulling for it. Each of those nodes has an independent probability of $k/n$ to hit a knowing node. Because $k \geq f$ we have that the probability that none of these nodes pulls from a node knowing about $\vec \mu$ is $(1 - k/n)^{n/2} < e^{-k} \leq e^{-f} = o(1)$. Lemma \ref{lem:knowledge-spreads} shows than that each node that does pull from a node that knows about $\vec \mu$ has a probability of $(1 - 1/q) = 1/2$ to learn $\vec \mu$. This means more generally we have at least $n/2$ nodes that have an independent chance of $k/2n$ to learn $\vec \mu$. For a small enough $\lambda$ it is clear that the probability that at least $\lambda k$ nodes learn about $\vec \mu$ can be made an arbitrarily small constant. 

In the second phase there are at least $n/2$ nodes that know about $\vec \mu$ and we want that of the remaining $k \leq n/2$ nodes at least a $\lambda$-fraction learns $\vec \mu$. Each of these nodes has a probability of at least $1/2 (1 - 1/q)$ to pull from a knowing node and learn $\vec \mu$ (see Lemma \ref{lem:knowledge-spreads}). Choosing $\lambda = 1/8$ suffices to guarantee that the probability that at least a $\lambda$-fraction learns $\vec \mu$ is at least $1/2$. The only reason that this probability can not be reduced is because if only one node remains to learn to learn about $\vec \mu$ a round is successful with probability exactly $1/2$.

Using the proof from Lemma \ref{lem:tail} it is easy to verify that choosing $t$ such that $t > k + O(\log n) \log t$ suffices to also make a union bound over these vectors at most $n^{-\Omega(1)}$. Combining this to a union bound over all vectors finished the proof by showing that the probability that after $t$ rounds not all vectors have spread is at most $n^{-\Omega(1)}$. 
\end{IEEEproof}\vspace{0.1cm}

\vspace{0.1cm} \begin{IEEEproof}[Proof of Lemma \ref{lem:exact-asynch-single}]
We want to show that running the protocol for $t = O(\frac{k}{\gamma} + T)$ rounds, where $T = O(\frac{\log^2 n}{\lambda})$ suffices to spread $k$ messages. Note that we always have $t>n$ and can also safely assume that $\log t = O(\log n)$. As a first step we define $p_i$ to be a lower bound for the probability that if $i$ nodes know about $\vec \mu$ in the next round one more node learns about $\vec \mu$. Note that by assumption and Lemma \ref{lem:knowledge-spreads} $p_i$ is lower bounded by $(1-1/q) \min\{i,n-i\} \lambda$ and $(1 - 1/q) \gamma$. We now look at $n$ phases in which we allow $\frac{\ln 3t}{p_i}$ tries for $i$ nodes informing the next node about $\vec \mu$. The number of rounds spend in successful phases sums up to at most $\sum={i=1}{n} \frac{\ln 3t}{p_i} \leq O(\frac{\log n}{\lambda}) \sum={i=1}{n/2} \frac{1}{i} \leq O(\frac{\log^2 n}{\lambda}) = T$. Lets now look at the probability that $\vec \mu$ has not spread after $t > T$ steps. In this case we have at least $t - T$ failures that can occur after any of the $n$ phases. The probability that at least $m$ errors occur after phase $i$ is at most $(1 - p_i)^{\frac{\ln 3t}{p_i} + m} < (3t)^{-1} (1 - \gamma/2)^m$. We thus get a $(2t)^{-1}$ factor for every phase that does not finish ``in time''. We also get a total factor of $(1 - \gamma/2)^{t-T}$ from all $t-T$ failures occurring after any round. Let $j$ be the number of phases that finish not ``in time''. There are exactly 
$\binom{(t-T) + j}{j} < (2t)^j$ ways of distributing the $t-T$ failures to these $j$ phases. Putting all this together we get the following upper bound on the probability that the algorithm did not converge after $t > T$ steps:
$$\sum_{j=1}^{n} (2t)^j (3t)^{-j} (1 - \gamma/2)^{t-T}  \leq \leq e^{- \gamma/2 (t-T)}$$
Choosing $t = O(\frac{k}{\gamma} + T)$ makes this smaller than $q^{-k} 2^{-n}$. Applying Theorem \ref{thm:reduction} now finishes the proof. 
\end{IEEEproof}\vspace{0.1cm}

\section*{Acknowledgments}
The author wants to thank Jon Kelner for his incredible help while finishing this write-up. He also wants to thank an anonymous reviewer of a related paper, David Karger and Muriel M\'{e}dard.


\bibliographystyle{IEEEtran}

\begin{thebibliography}{10}
\providecommand{\url}[1]{#1}
\csname url@samestyle\endcsname
\providecommand{\newblock}{\relax}
\providecommand{\bibinfo}[2]{#2}
\providecommand{\BIBentrySTDinterwordspacing}{\spaceskip=0pt\relax}
\providecommand{\BIBentryALTinterwordstretchfactor}{4}
\providecommand{\BIBentryALTinterwordspacing}{\spaceskip=\fontdimen2\font plus
\BIBentryALTinterwordstretchfactor\fontdimen3\font minus
  \fontdimen4\font\relax}
\providecommand{\BIBforeignlanguage}[2]{{%
\expandafter\ifx\csname l@#1\endcsname\relax
\typeout{** WARNING: IEEEtran.bst: No hyphenation pattern has been}%
\typeout{** loaded for the language `#1'. Using the pattern for}%
\typeout{** the default language instead.}%
\else
\language=\csname l@#1\endcsname
\fi
#2}}
\providecommand{\BIBdecl}{\relax}
\BIBdecl

\bibitem{algebraicgossip-deb-med-choute-06-transinf}
S.~Deb, M.~Medard, and C.~Choute, ``Algebraic gossip: a network coding approach
  to optimal multiple rumor mongering,'' \emph{IEEE Transactions on Information
  Theory}, vol.~52, no.~6, pp. 2486 -- 2507, 2006.

\bibitem{informationdissemination05}
------, ``On random network coding based information dissemination,'' in
  \emph{Proceedings of the International Symposium on Information Theory
  (ISIT)}, 2005, pp. 278 --282.

\bibitem{mosk2006information}
D.~Mosk-Aoyama and D.~Shah, ``{Information dissemination via network coding},''
  in \emph{Proceedings of the IEEE International Symposium on Information
  Theory (ISIT)}, 2006, pp. 1748--1752.

\bibitem{vasudevan2009algebraic}
D.~Vasudevan and S.~Kudekar, ``{Algebraic gossip on Arbitrary Networks},''
  \emph{Arxiv preprint arXiv:0901.1444}, 2009.

\bibitem{borokhovich2010tight}
M.~Borokhovich, C.~Avin, and Z.~Lotker, ``{Tight Bounds for Algebraic Gossip on
  Graphs},'' \emph{Arxiv preprint arXiv:1001.3265 v1 [cs. IT]}, 2010.

\bibitem{gkantsidis2005network}
C.~Gkantsidis and P.~Rodriguez, ``{Network coding for large scale content
  distribution},'' in \emph{Proceedings of the 24th International Conference on
  Computer Communications (INFOCOM)}, vol.~4, 2005.

\bibitem{adhocrouting}
Z.~J. Haas, J.~Y. Halpern, and L.~Li, ``Gossip-based ad hoc routing,''
  \emph{IEEE/ACM Transactions on Networks (TON)}, vol.~14, no.~3, pp. 479--491,
  2006.

\bibitem{demers1987epidemic}
A.~Demers, D.~Greene, C.~Hauser, W.~Irish, J.~Larson, S.~Shenker, H.~Sturgis,
  D.~Swinehart, and D.~Terry, ``{Epidemic algorithms for replicated database
  maintenance},'' in \emph{Proceedings of the 6th Symposium on Principles of
  Distributed Computing (PODC)}, 1987, pp. 1--12.

\bibitem{epidemicdatabase}
D.~Agrawal, A.~El~Abbadi, and R.~C. Steinke, ``Epidemic algorithms in
  replicated databases (extended abstract),'' in \emph{Proceedings of the 16th
  Symposium on Principles of Database Systems (PODS)}, 1997, pp. 161--172.

\bibitem{kempe2004spatial}
D.~Kempe, J.~Kleinberg, and A.~Demers, ``{Spatial gossip and resource location
  protocols},'' \emph{Journal of the ACM (JACM)}, vol.~51, no.~6, pp. 943--967,
  2004.

\bibitem{hedetniemi88}
S.~M. Hedetniemi, S.~T. Hedetniemi, and A.~L. Liestman, ``A survey of gossiping
  and broadcasting in communication networks,'' \emph{Networks}, vol.~18, pp.
  319--349, 1988.

\bibitem{topkis85}
D.~M. Topkis, ``Concurrent broadcast for information dissemination,''
  \emph{IEEE Transactions on Software Engineering}, vol. SE-11, no.~10, 1985.

\bibitem{aspnes09}
J.~Aspnes and E.~Ruppert, ``An introduction to population protocols,'' in
  \emph{Middleware for Network Eccentric and Mobile Applications},
  B.~Garbinato, H.~Miranda, and L.~Rodrigues, Eds.\hskip 1em plus 0.5em minus
  0.4em\relax Springer-Verlag, 2009, pp. 97--120.

\bibitem{hromkovic96}
J.~Hromkovi\v{c}, R.~Klasing, B.~Monien, and R.~Peine, ``Dissemination of
  information in interconnection networks (broadcasting \& gossiping),''
  \emph{Combinatorial Network Theory}, pp. 125--212, 1996.

\bibitem{kempe03}
D.~Kempe, A.~Dobra, and J.~Gehrke, ``Gossip-based computation of aggregate
  information,'' in \emph{Proceedings of 44th Symposium on Foundations of
  Computer Science (FOCS)}, 2003, pp. 482--491.

\bibitem{kempe02}
D.~Kempe and J.~Kleinberg, ``Protocols and impossibility results for
  gossip-based communication mechanisms,'' in \emph{Proceedings of 43rd
  Symposium on Foundations of Computer Science (FOCS)}, 2002, pp. 471--480.

\bibitem{karp00}
R.~Karp, C.~Schindelhauer, S.~Shenker, and B.~V\"ocking, ``Randomized rumor
  spreading,'' in \emph{Proceedings of 41st Symposium on Foundations of
  Computer Science (FOCS)}, 2000, pp. 565--574.

\bibitem{minski-spreading}
Y.~Minski, ``{Spreading rumors cheaply, quickly, and reliably, 2002},'' Ph.D.
  dissertation, Ph. D. Thesis, Cornell University.

\bibitem{spreadingwithconductance}
F.~Chierichetti, S.~Lattanzi, and A.~Panconesi, ``{Almost tight bounds for
  rumour spreading with conductance},'' in \emph{Proceedings of the 42nd ACM
  Symposium on Theory of Computing (STOC)}, 2010, pp. 399--408.

\bibitem{ahlswede2000network}
R.~Ahlswede, N.~Cai, S.~Li, and R.~Yeung, ``{Network information flow},''
  \emph{IEEE Transactions on Information Theory}, vol.~46, no.~4, pp.
  1204--1216, 2000.

\bibitem{li2003linear}
S.~Li, R.~Yeung, and N.~Cai, ``{Linear network coding},'' \emph{IEEE
  Transactions on Information Theory}, vol.~49, no.~2, pp. 371--381, 2003.

\bibitem{ho2003benefits}
T.~Ho, R.~Koetter, M.~Medard, D.~Karger, and M.~Effros, ``{The benefits of
  coding over routing in a randomized setting},'' in \emph{Proceedings of the
  IEEE International Symposium on Information Theory (ISIT}, 2003, pp.
  442--442.

\bibitem{algebraicgossip-deb-medard04-allerton}
S.~Deb and M.~M\'{e}dard, ``{Algebraic gossip: a network coding approach to
  optimal multiple rumor mongering},'' in \emph{Proceedings 42rd Allerton
  Conference on Communication, Control, and Computing}, 2004.

\bibitem{katti2005importance}
S.~Katti, D.~Katabi, W.~Hu, H.~Rahul, and M.~Medard, ``{The importance of being
  opportunistic: Practical network coding for wireless environments},'' in
  \emph{Proceedings 43rd Allerton Conference on Communication, Control, and
  Computing}, 2005.

\bibitem{chou2003practical}
P.~Chou, Y.~Wu, and K.~Jain, ``{Practical network coding},'' in
  \emph{Proceedings of the 41st Allerton Conference on Communication Control
  and Computing}, vol.~41, no.~1, 2003, pp. 40--49.

\bibitem{efficientbroadcastingusingNC}
C.~Fragouli, J.~Widmer, and J.-Y. Le~Boudec, ``Efficient broadcasting using
  network coding,'' \emph{IEEE/ACM Transactions Netw.}, vol.~16, no.~2, pp.
  450--463, 2008.

\bibitem{katti2008xors}
S.~Katti, H.~Rahul, W.~Hu, D.~Katabi, M.~M{\'e}dard, and J.~Crowcroft, ``{XORs
  in the air: practical wireless network coding},'' \emph{IEEE/ACM Transactions
  on Networking (TON)}, vol.~16, no.~3, pp. 497--510, 2008.

\bibitem{fragouli2006network}
C.~Fragouli, J.~Widmer, and J.~Boudec, ``{A network coding approach to energy
  efficient broadcasting: from theory to practice},'' in \emph{Proceedings of
  the 25th International Conference on Computer Communications (INFOCOM)},
  2006.

\bibitem{KLO}
F.~Kuhn, N.~Lynch, and R.~Oshman, ``Distributed computation in dynamic
  networks,'' in \emph{Proceedings of the 42nd Symposium on Theory of Computing
  (STOC)}, 2010, pp. 557--570.

\bibitem{afek95}
Y.~Afek and D.~Hendler, ``On the complexity of gloabl computation in the
  presence of link failures: The general case,'' \emph{Distributed Computing},
  vol.~8, no.~3, pp. 115--120, 1995.

\bibitem{afek87}
Y.~Afek, B.~Awerbuch, and E.~Gafni, ``Applying static network protocols to
  dynamic networks,'' in \emph{Proceedings\ of 28th Symposium\ on Foundations
  of Computer Science (FOCS)}, 1987, pp. 358--370.

\bibitem{awerbuch88}
B.~Awerbuch and M.~Sipser, ``Dynamic networks are as fast as static networks,''
  in \emph{Proceedings\ of 29th Symposium\ on Foundations of Computer Science
  (FOCS)}, 1988, pp. 206--220.

\bibitem{awerbuch92}
B.~Awerbuch, B.~Patt-Shamir, D.~Peleg, and M.~E. Saks, ``Adapting to
  asynchronous dynamic networks,'' in \emph{Proceedings\ of the 24th Symposium
  on Theory of Computing (STOC)}, 1992, pp. 557--570.

\bibitem{dijkstra74}
E.~Dijkstra, ``Self-stabilizing systems in spite of distributed control,''
  \emph{Communications of the ACM}, vol.~11, pp. 643--644, 1974.

\bibitem{mosk-aoyama06}
D.~Mosk-Aoyama and D.~Shah, ``Computing separable functions via gossip,'' in
  \emph{Proceedings of 25th Symposium on Principles of Distributed Computing
  (PODC)}, 2006, pp. 113--122.

\bibitem{bar-yehuda92}
R.~Bar-Yehuda, O.~Goldreich, and A.~Itai, ``On the time complexity of broadcast
  in radio networks: An exponential gap between determinism and
  randomization,'' \emph{Journal of Computer and System Sciences (JCSS)},
  vol.~45, no.~1, pp. 104--126, 1992.

\bibitem{clementi01}
A.~E.~G. Clementi, A.~Monti, and R.~Silvestri, ``Distributed multi-broadcast in
  unknown radio networks,'' in \emph{Proceedings\ of 20th Symposium\ on
  Principles of Distributed Computing (PODC)}, 2001, pp. 255--263.

\bibitem{clementi09}
A.~E.~F. Clementi, A.~Monti, F.~Pasquale, and R.~Silvestri, ``Broadcasting in
  dynamic radio networks,'' \emph{Journal of Computer and System Sciences
  (JCSS)}, vol.~75, no.~4, pp. 213--230, 2009.

\bibitem{baumann09}
H.~Baumann, P.~Crescenzi, and P.~Fraigniaud, ``Parsimonious flooding in dynamic
  graphs,'' in \emph{Proceedings\ of 28th Symposium\ on Principles of
  Distributed Computing (PODC)}, 2009, pp. 260--269.

\end{thebibliography}

\end{document}